\documentclass[aps,prb,twocolumn,amsmath]{revtex4}
\usepackage{color,graphicx}
\usepackage{bm}
\usepackage{amsmath}
\usepackage{amsfonts}
\usepackage{amssymb}

\newcommand{\be}{\begin{equation}}
\newcommand{\ee}{\end{equation}}
\newcommand{\bea}{\begin{eqnarray}}
\newcommand{\eea}{\end{eqnarray}}
\newcommand{\bei}{\begin{itemize}}
\newcommand{\eei}{\end{itemize}}

\newcommand{\br}{{\bf{r}}}
\newcommand{\bk}{{\bf{k}}}

\begin{document}

\title{Condensed Groundstates of Frustrated Bose-Hubbard Models}

\author{G. M\"{o}ller and N. R. Cooper}
\affiliation{Theory of Condensed Matter Group, Cavendish Laboratory, J.~J.~Thomson Ave., Cambridge CB3~0HE, UK}

\begin{abstract}
  We study theoretically the groundstates of two-dimensional Bose-Hubbard
  models which are frustrated by gauge fields. Motivated by  recent proposals
  for the implementation of optically induced gauge potentials,
  we focus on the situation in which the imposed gauge fields give rise to a
  pattern of staggered fluxes, of magnitude $\alpha$ and alternating in sign
  along one of the principal axes. For $\alpha=1/2$ this model is equivalent
  to the case of uniform flux per plaquette $n_\phi=1/2$, which, in the
  hard-core limit, realizes the ``fully frustrated'' spin-1/2 XY model.  We
  show that the mean-field groundstates of this frustrated Bose-Hubbard model
  typically break translational symmetry.  We introduce a general
  numerical technique to detect broken symmetry condensates in exact
  diagonalization studies.  Using this technique we show that, for all cases
  studied, the groundstate of the Bose-Hubbard model with staggered flux
  $\alpha$ is condensed, and we obtain quantitative determinations of the
  condensate fraction. We discuss the experimental consequences of our
  results.  In particular, we explain the meaning of gauge-invariance in
  ultracold atom systems subject to optically induced gauge potentials, and show
  how the ability to imprint phase patterns prior to expansion can allow very
  useful additional information to be extracted from expansion images.
\end{abstract}
\date{September 22, 2010}
\pacs{
}
\maketitle

\section{Introduction}

One of the most striking aspects of the physics of Bose-Einstein condensed
systems is their response to rotation. The rotation plays the role of a
uniform magnetic field, which frustrates the uniform condensate, forcing it into a state containing quantized vortices and carrying non-vanishing
currents.\cite{blochdz,fetterreview,advances} Theory shows that at
sufficiently high vortex density this frustration can lead to the breakdown of
Bose-Einstein condensation, and the formation of a series of strongly
correlated quantum phases which can be viewed as bosonic analogues of the
fractional quantum Hall states.\cite{advances}

In typical magnetically trapped Bose gases\cite{SchweikhardCEMC92} practical
limitations on the rotation rate (vortex density) are such that strongly
correlated phases are expected only at a very low particle density where the interaction energy scale is very small.\cite{advances} As a
result, it has proved difficult to reach this strongly correlated
regime. (However, see Ref.~\onlinecite{gemelke} for interesting recent results
for systems with small particle numbers.)

It has been proposed that one can exploit the strong interactions that are
available in systems of bosonic atoms confined to optical
lattices\cite{blochdz} to enhance the possibility of achieving these
correlated phases. In this context, the natural model to consider is the
Bose-Hubbard model with uniform effective magnetic flux
[Eq.~(\ref{eq:hamiltonian})]. This ``frustrated'' Bose-Hubbard model can show
very interesting physics, far beyond the physics of the usual Bose-Hubbard
model.\cite{fisherbh}  Atomic systems well-described by this frustrated
Bose-Hubbard have been studied experimentally by using rotating optical
lattices,\cite{tung,footlattice} albeit so far limited to situations of large
lattice constants and large numbers of particles per lattice site which are
outside the strongly correlated regime.  However, a series of
theoretical
proposals\cite{JakschZoller,mueller,sorensen:086803,palmer,palmer:013609,hafezi-2007,gerbier}
indicate that it should be possible to imprint strong gauge fields on an optical lattice, and thereby realize a regime where interactions are
strong, and with both the particle number per site, $n$, and vortex number per
plaquette, $n_\phi$, of order one. In this regime, theory shows that there are
strongly correlated phases representative of the continuum quantum Hall states
limit\cite{sorensen:086803,hafezi-2007,hafezi-epl} as well as related
interesting strongly correlated phases that are stabilized by the lattice
itself.\cite{mollercooper-cf} Other candidates are related to Mott physics.\cite{Umucalilar:2010p395,Powell:2010p303}

Our confidence in the existence of strongly correlated phases of the
frustrated Bose-Hubbard model relies on the results of large-scale numerical
exact diagonalization
studies.\cite{sorensen:086803,hafezi-2007,hafezi-epl,mollercooper-cf}
However, these studies have found evidence for strongly correlated phases only
in a relatively small region of parameter space (spanned by the particle density
per site $n$, flux per plaquette $n_\phi$, and interaction strength
$U/J$). There are surely competing {\it condensed} phases, which can be viewed
as vortex lattices that are pinned by the lattice.\cite{duriclee} An important question emerges
from the point of view of these numerical approaches: How does one determine
condensation
in exact diagonalization studies?  In conventional condensed systems, one
looks for the maximum eigenvalue of the single particle density matrix of the
groundstate.\cite{yang}  However, here the condensed states are (pinned)
vortex lattices, and therefore break translational symmetry. As a result, one
expects a degeneracy of the spectrum in the thermodynamic
limit.\cite{cwg,advances} How does one quantify the degree of condensation?

In this paper we propose a powerful general numerical method that can be used
to identify and characterize condensed groundstates which break a symmetry of
the Hamiltonian. We use this to study several cases of interest in the context
of optically induced gauge potentials.\cite{dalibardreview} Optically induced
gauge potentials have recently been implemented experimentally without an
optical lattice.\cite{lin:130401,spielmanfield} These successes encourage a
high degree of optimism that the related schemes on optical
lattices\cite{JakschZoller,gerbier} will also be successful.

Motivated by the proposals of Jaksch and Zoller\cite{JakschZoller} and Gerbier
and Dalibard,\cite{gerbier} in this paper we focus not on the case of a
uniform magnetic field, but on the case of a two-dimensional square lattice
with a {\it staggered} magnetic field, with a flux per plaquette of magnitude
$\alpha$ that alternates in sign along one of the principal axes. This flux
configuration involves much a simpler experimental implementation than the
case of uniform flux. As described below, for the special case of $\alpha=1/2$
this is equivalent to the uniform flux.  In this case, the model simulates a
quantum version of the ``fully frustrated'' XY model. For other values of
$\alpha$, it represents a class of frustrated quantum spin models. Related
but different staggered flux Hamiltonians can be generated by time-dependent
lattice potentials as discussed in Ref.~\onlinecite{morais}.

Based on numerical exact diagonalizations, we provide evidence
showing that the groundstate breaks translational invariance, and is condensed
for all flux densities and (repulsive) interaction strengths. We evaluate the
condensate fraction and condensate wavefunctions from the exact
diagonalization results.  We describe how evidence for translational symmetry
breaking can be found in measurements of the real-space and momentum space
(expansion) profiles, and how these can be used to determine the condensate
depletion.  As part of this work, we explain some important general aspects of
expansion imaging of systems involving optically induced gauge potentials.

\section{Model}

We shall study the properties of the two-dimensional Bose-Hubbard model
subject to an abelian gauge potential, as described by the Hamiltonian
\bea
\nonumber
\hat{H} & = & -J \sum_{\langle i,j\rangle } \left[\hat{a}_i^\dag\hat{a}_j e^{i A_{ij}} +
    \hat{a}_j^\dag\hat{a}_i e^{i A_{ji}}\right] + \frac{U}{2}\sum_i \hat{n}_i(\hat{n}_i-1) 
%\\
% & & - \mu\sum_i \hat{n}_i
\label{eq:hamiltonian}
\eea
The operator $\hat{a}_i^{(\dag)}$ destroys (creates) a boson on the lattice
site $i$, which we choose to form a square lattice; $U$ describes the onsite
repulsion ($U\geq 0$ is assumed throughout); $J$ is the nearest-neighbour
tunneling energy. The Hamiltonian conserves the total number of bosons,
$\hat{N} = \sum_i \hat{n}_i = \sum_i \hat{a}^\dag_i\hat{a}_i$. Throughout this
work, we shall consider the system to be uniform, with $N$ chosen such that
the mean particle density per lattice site is $n$. The results of these
studies can be used within the local density approximation to model
experimental systems which have an additional trapping potential.

\begin{figure}
\includegraphics[width=0.8\columnwidth]{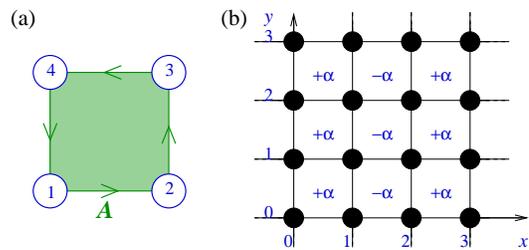} 
\caption{(a) The gauge-invariant flux through the plaquette, $n_\phi$, is
defined by $n_\phi \equiv \frac{1}{2\pi} \sum_{i,j} A_{ij} = \frac{1}{2\pi}\left(A_{12}+A_{23}+A_{34}+A_{41}\right)$
(b) The simplest optically induced gauge potential to imprint on the square lattice\protect\cite{JakschZoller,gerbier} has an alternating pattern of fluxes of magnitude $\alpha$, Eq.~(\ref{eq:staggered}).}
\label{fig:plaquette}
\end{figure}

The fields $A_{ij}$ (which satisfy $A_{ij} = -A_{ji}$) describe the imposed
gauge potential.  All of the physics of the system defined by the Hamiltonian
(\ref{eq:hamiltonian}) (energy spectrum, response functions, etc.)  is
gauge-invariant. Therefore, its properties depend only on the fluxes through plaquettes
\be n^{a}_\phi \equiv \frac{1}{2\pi} \sum_{i,j\in a}
A_{ij} 
\label{eq:flux}
\ee
where $a$ labels the plaquette, and the sum represents the
directed sum of the gauge fields around that plaquette (the discrete
version of the line integral), as illustrated in
Fig.~\ref{fig:plaquette}(a).
Since each phase $A_{ij}$ is defined modulo $2\pi$, the gauge
invariant fluxes (\ref{eq:flux}) are defined modulo $1$ (i.e.  are invariant under
$n_\phi^a \to n_\phi^a +1$), so they can be restricted to
the interval $-1/2< n^a_\phi \leq 1/2$.

The gauge-invariant fluxes through the plaquettes lead to an intrinsic
``frustration'' of condensed (superfluid) phases on the Bose-Hubbard
system. This is best understood in the case of strong interactions, $U
\gg J$, where double occupancy is excluded.  In this hard-core limit,
the Bose-Hubbard model is equivalent to a spin-1/2 quantum magnet,
using the standard mapping $\hat{s}_i^z = \hat{n}_i-\frac{1}{2}$,
$\hat{s}_i^+ = \hat{a}_i^\dag$, $\hat{s}_i^- = \hat{a}_i$, with
Hamiltonian (up to a constant shift in energy)
\bea
%\nonumber
\hat{H}_{\rm h-c} & = & -J \sum_{\langle i,j\rangle} \left[\hat{s}_i^+\hat{s}^-_j e^{i A_{ij}} +
   \hat{s}_j^+\hat{s}^-_i e^{-i A_{ij}} \right]  
%- \mu\sum_i \hat{s}^z_i 
\label{eq:spham}
\eea
(The
conservation of particle number becomes conservation of $\hat{S}^z =
\sum_i \hat{s}^z_i = \hat{N} -1/2$.)
This Hamiltonian describes a quantum spin-1/2 magnet, experiencing XY
nearest neighbour spin exchange interactions. These exchange
interactions are {\it ``frustrated"} by the gauge fields.  
The ``frustration'' can be seen by considering the natural mean-field limit of
the spin Hamiltonian (\ref{eq:spham}), generalizing from spin-$1/2$ to
spin-$S$ and taking the $S\to \infty$ limit.\cite{duriclee} Then, the (vector
of) spin operators $\hat{s}_i$ can be replaced by the classical vector
$\vec{s}$ of fixed length $S$. It is convenient to parameterize this vector as
\be \vec{s} = S(\sin\theta \cos\phi,\sin\theta
\cos\phi,\cos\theta) 
\label{eq:means}
\ee
which represents the spin by the polar and azimuthal angles $\theta,\phi$.
The Hamiltonian becomes the
(classical) energy functional
\be
H_{\rm mft} = -2JS^2 \sum_{\langle i,j\rangle} \sin\theta_i\sin\theta_j
    \cos(\phi_i-\phi_j+ A_{ij})
% - \mu S\sum_i\cos\theta_i
\label{eq:sphammft}
\ee
 and it is natural define the fractional occupation of the lattice
sites by $n_i = (1/2) (1+ \cos\theta_i)$.

\begin{figure}
\includegraphics[width=0.8\columnwidth]{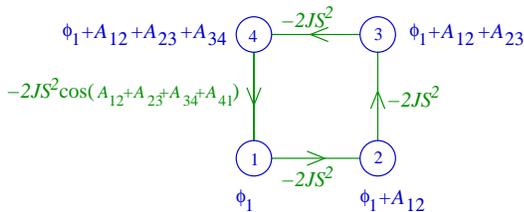}$\qquad\;$
\caption{Illustration of the frustrated coupling of the XY spin
  model around a plaquette with non-zero flux $n_\phi$.  For a given
  phase $\phi_1$ at site 1, one can choose $\phi_2$, $\phi_3$, and
  $\phi_4$ to maximally reduce the energy on three of the bonds
  (with a contribution of $-JS^2$ to the energy for each).  However, the exchange energy on the remaining
  bond is $ -JS^2\cos(A_{12} + A_{23} + A_{34} + A_{41}) =
  -JS^2\cos(2\pi n_\phi)$.  For $n_\phi\neq 0$, this bond cannot also 
be fully satisfied, indicating the magnetic frustration. Maximal
  frustration occurs for $n_\phi =1/2$.}
\label{fig:choices}
\end{figure}

The restricted space of configurations with $\theta_i=\pi/2$ is
important for groundstate configurations with (uniform) density $n =
1/2$. In this case, $s_z =0$, so all spins lie in the $x$-$y$-plane and
the Hamiltonian becomes that of the frustrated XY
model.\cite{teitel} Here, ``frustration'' refers to the fact that,
with $n_\phi \neq 0$ for any plaquette, the angles $\phi_i$ around
this plaquette cannot be chosen in order to maximally satisfy the XY
exchange couplings. This is illustrated in Fig.~\ref{fig:choices}.

Maximal frustration occurs for $n_\phi = 1/2$. This situation is referred to
as the ``fully frustrated'' case.  The ``fully frustrated'' classical XY
model has been much studied as an interesting frustrated classical magnet with a
non-standard thermal phase transition.\cite{villain} The frustrated
Bose-Hubbard model that we study here is a class of frustrated spin-$1/2$
quantum systems that are analogous to this frustrated classical model. We
shall focus on the nature of the groundstates of the system.

\section{Staggered Fluxes}

Motivated by proposals for optically induced gauge
potentials,\cite{JakschZoller,gerbier} we consider a situation in
which the flux is staggered in the $x$ direction.  Specifically, we
choose the gauge (and geometry) described in
Ref.~\onlinecite{gerbier}, for which
\be
A_{ij} = 2\pi \alpha \, y_i
(x_i-x_j)\times(-1)^{x_i}
\label{eq:gauge}
\ee
where $(x_i,y_i)$ is the pair of integers that define the position of
lattice site $i$ ({\it i.e.} the cartesian co-ordinates position in
units of the lattice constants, $a_x$, $a_y$, in the $x$- and
$y$-directions, as shown in Fig.\ref{fig:plaquette}(b)). Since the lattice is square with nearest-neighbour
hopping, this has the effect that hopping in the $y$ direction has no
gauge field ($A_{ij}=0$) while hopping in the $x$ direction involves a
phase $A_{ij} = 2\pi \alpha y_i (-1)^{x_i}$ that alternates from one
row to the next.  The magnitude of the flux per plaquette is
\be
n^{a}_\phi = (-1)^{x_a} \alpha \,,
\label{eq:staggered}
\ee
where the position $x_a$ is defined by the $x$-position of the
``bottom-left'' corner of the plaquette ({\it i.e.}  the minimum $x_i$ for
all sites $i$ surrounding the plaquette $a$).  The flux is staggered
in the $x$-direction, as shown in Fig.~\ref{fig:plaquette}(b).

A special situation arises for $\alpha=1/2$. Then, the case of alternating
fluxes of $n^a_\phi = (-1)^{x_a}\frac{1}{2}$ is gauge-equivalent to that of 
uniform flux $n^a_\phi = \frac{1}{2}$. The (gauge-invariant)
properties of this case have higher translational symmetry than the
case of $\alpha\neq 0,1/2$. Furthermore, a gauge can be chosen in
which $e^{i A_{ij}}$ is real, meaning that the Hamiltonian is
time-reversal symmetric. Indeed the gauge (\ref{eq:gauge}) has this
property.
While gauge invariance allows the physics of (\ref{eq:hamiltonian}) to
be studied in any gauge, as we shall describe below, the {\it
  expansion images} of the atomic gas are gauge-dependent.  We shall
therefore be clear to specify the gauge considered.

Under the conditions that we study, significant insight into the
physics of the frustrated Bose-Hubbard model can be obtained by studying
its properties within mean-field theory. Indeed, one important goal of
this work is to show how the mean-field groundstates emerge from the
results of exact diagonalization studies.

\subsection{Single Particle Spectrum}

We first ignore interactions, and study the single particle energy
eigenstates. For the gauge field we consider (\ref{eq:gauge}), the unit cell
can be chosen to have size $2\times 1$ (in the $x$- and $y$-directions), and
the energy eigenstates are then
\be
\psi_i = e^{i (k_x x_i + k_y y_i)} \times \left\{ \begin{array}{l}
\psi_e \quad\quad\quad\quad x_i \;\mbox{even}\\
\psi_o e^{i 2\pi\alpha y_i} \quad x_i \;\mbox{odd}
\end{array}
\right.
\ee
with the momenta in the ranges $-\pi/2\leq k_x < \pi/2$ and $-\pi\leq k_y <
\pi$. [We express $k_x$ and $k_y$ in units of $1/a_x$ and $1/a_y$
respectively.]
The energy eigenvalues $E$ and eigenfunctions
within the unit cell, $(\psi_e,\psi_o)$,
 follow from
\be
-2J \left(\begin{array}{cc}
\cos(k_y) & \cos(k_x) \\
\cos(k_x) & \cos(k_y+2\pi\alpha) 
\end{array}\right)
\left(\begin{array}{c}
\psi_e\\
\psi_o
\end{array}\right)
= E
\left(\begin{array}{c}
\psi_e\\
\psi_o
\end{array}\right)
\label{eq:matrix}
\ee

\begin{figure}
\includegraphics[width=0.9\columnwidth]{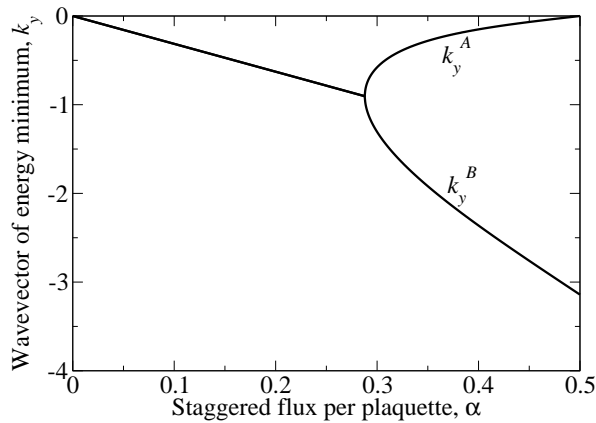}
\caption{The single-particle groundstate is non-degenerate for $\alpha< 0.289$, but is two-fold degenerate for $\alpha > 0.289$. The two minima are at wavevectors $k^A_y, k^B_y$ shown above (both have $k_x=0$).
}
\label{fig:minima}
\end{figure}

The lowest energy state has $k_x=0$. For $\alpha <
\frac{1}{\pi}\arccos\left((\sqrt{5}-1)/2\right) \simeq 0.288$, there
is a single minimum at $k_y=\pi\alpha$. However, for $\alpha > 0.288$
this minimum splits in two, and there are two {\it degenerate} minima
defining single particle states $\psi^A_i$ and $\psi^B_i$.  The
wavevectors of these states, $k_y^{A,B}$ are shown in
Fig.~\ref{fig:minima}. They are related by $k^A_y+k_y^B = -2\pi
\alpha$.\footnote{Note that the positions $k^{A/B}$ of these minima are gauge dependent.}  
Hence $\cos(k^{A/B}_y) = \cos(k^{B/A}_y+2\pi\alpha)$, so the
Hamiltonian (\ref{eq:matrix}) is the same for both states provided the
two sites of the unit cell are swapped. Therefore, $\psi^A$ and $\psi^B$
are exactly degenerate, and their wavefunctions related by
$\psi^{A}_{e/o} = \psi^{B}_{o/e}$.

The appearance of two degenerate minima in the single-particle spectrum leads
to the question of whether the groundstate is a ``simple'' condensate, or
``fragmented''.\cite{mueller:033612} Following the general
result,\cite{nozieresfragment} one expects that weak repulsive interactions
will lead to a simple condensate in which all particles condense the same
single particle state.  The nature of this condensate can depend on the
properties of the two states, and the nature of the interactions. However, if
the condensate wavefunction contains non-zero weights of both states, with
wavevectors $k_y^A$ and $k_y^B$, then the condensate breaks
translational symmetry in the $y$ direction.

Similar physics -- of a two-component Bose gas -- has been discussed recently
in the context of optically induced gauge potentials in the
continuum.\cite{spielmannonab,hononabelian,Wang:arXiv1006.514} For the case we consider, 
one principle difference is
that there is an underlying lattice periodicity. Phases with broken
translational symmetry can therefore ``lock'' to the lattice periodicity,
leading to condensed groundstates with a broken {\it discrete} translational
symmetry. Furthermore, the unit cells of the phases that we describe contain
particle currents, which can be viewed as arrays of vortices and antivortices.

In the following we shall explore the nature of this symmetry
breaking. A principle result will be to show that the results of exact
diagonalization studies are consistent with simple condensation in a
symmetry broken state. We shall focus on two cases:

(i) $\alpha =1/2$. Here, the model is equivalent to the case of
uniform fluxes $n_\phi^a = 1/2$. The (gauge-invariant) system
therefore enjoys the full translational symmetry of the lattice.  The
minima are spaced by $\Delta k_y = \pi$, suggesting that the broken
symmetry state will have translational period $2$ in the
$y$-direction. The symmetry broken state will have two
degenerate configurations.

(ii) $\alpha =
\frac{1}{\pi}\arccos\left(\frac{1}{2}\sqrt{\frac{1}{2}(9-\sqrt{65})}\right)
\simeq 0.389$. This value is chosen such that $\Delta k_y = 2\pi/3$,
such that the broken translational symmetry state has period $3$ in
the $y$-direction. The symmetry broken state will have three
degenerate configurations.

Throughout this work, we concentrate on cases where the particle density $n$
is non-integer. Thus, we do not discuss the properties of the Mott insulator
states. While frustration by the gauge fields can affect the stability of the
Mott insulating states,\cite{oktel,goldbaummueller} the qualitative form of these
insulating states is unchanged. We focus on the nature of the superfluid
states, the properties of which are much changed in the presence of the
frustrating gauge fields.

\subsection{Gross-Pitaevskii Theory}

One can understand the effects of weak interactions, {\it i.e.} $U\ll nJ$, within
Gross-Pitaevskii mean-field theory. One assumes complete condensation,
forming a many-body state
\be |\Psi_c\rangle \equiv \left( \sum_i \psi^c_i
  \hat{a}^\dag_i\right)^N |0\rangle 
\label{eq:condensedstate}
\ee
where $\psi^c_i$ is the (normalized) condensate wavefunction.  The
condensate wavefunction is
determined by viewing (\ref{eq:condensedstate})
as a variational state, and
minimizing the average energy per particle
\bea
\label{eq:ke}
\frac{\langle \hat{H}\rangle}{N}  & = & -J  \sum_{\langle i,j\rangle } \left[\psi^{c*}_i\psi^c_j e^{i A_{ij}} +\psi^{c*}_j\psi^c_i e^{i A_{ji}}\right] \\
 & & + \frac{U}{2}(N-1)\sum_i |\psi^c_i|^4 - \frac{U}{2N}
\label{eq:int}
\eea 
For large systems $N\gg 1$ the typical interaction energy
(\ref{eq:int}) is of the order of $Un$, while the kinetic energy
(\ref{eq:ke}) is of order $J$.  

In the weak coupling limit, $nU \ll J$, the groundstates of the
Gross-Pitaevskii equation can be obtained by assuming that the
condensate consists only of those states that minimize the
single-particle kinetic energy. For $\alpha < \alpha_c $, there is a
single minimum, so the condensate will be formed from this state
alone.  For $\alpha > \alpha_c$, there are two degenerate minima, and
we should write
\be
\psi^c = A \psi_{k_A} + B \psi_{k_B}
\label{eq:linear}
\ee
which (schematically) denotes a linear superposition of the states in these
two minima.  Then one chooses $A$ and $B$ to minimize the interaction energy
$\sum_i |\psi^c_i|^4$.

\subsubsection{The case $\alpha = 1/2$}
\label{sec:alpha_0.5}
\begin{figure}
\includegraphics[width=0.99\columnwidth]{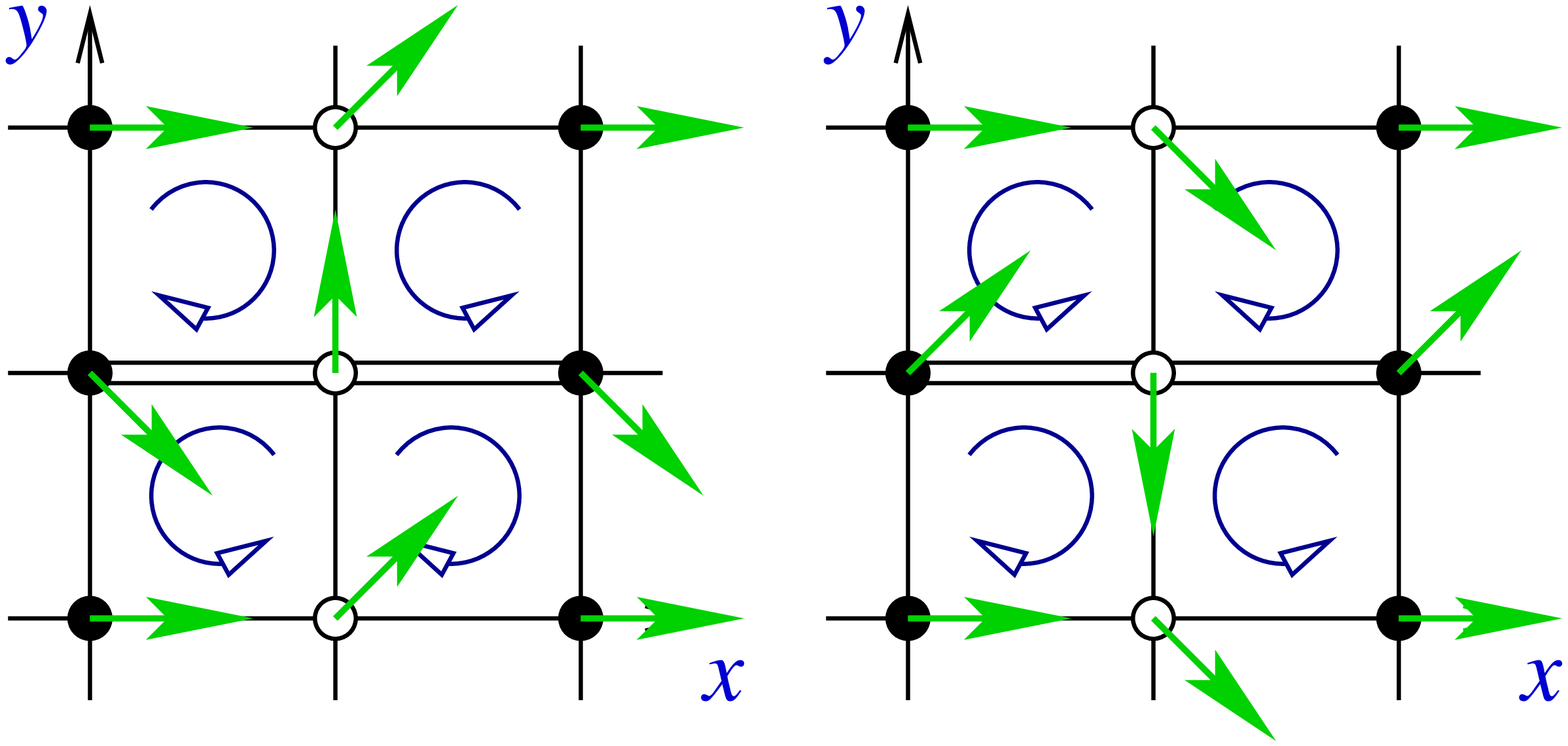}
\caption{Illustration of the 
two (degenerate)
  mean-field groundstates of the fully frustrated model, $n_\phi =
  1/2$. The arrows on each lattice site represent the phase of the condensate 
wavefunction (the amplitude is constant). The two phases are characterized by
staggered circulating currents, in directions illustrated within the plaquettes.
Double lines mark links where a `phase' $e^{i\pi}$ is imprinted.
}
\label{fig:villain}
\end{figure}

This is the case of the fully frustrated Bose-Hubbard model, $\alpha=1/2$,
which is gauge equivalent to uniform plaquette fluxes of $n_\phi = 1/2$.
Minimizing the interaction energy within states of the form (\ref{eq:linear})
leads to two solutions which we can denote schematically by
\be
\psi^c_\pm  = \frac{1}{\sqrt{2}}\left[ \psi_{k_A} \pm i  \psi_{k_B}\right]
\label{eq:staggeredstates}
\ee
In detail, the wavefunctions are
\be
\psi^c_{\pm,i} = \frac{1}{2\sqrt{2-\sqrt{2}}} \times 
\left\{ \begin{array}{l}
1\pm i(\sqrt{2}-1)(-1)^{y_i}  \quad\quad x_i \;\mbox{even}\\
(\sqrt{2}-1)(-1)^{y_i} \pm i \quad\quad\quad x_i \;\mbox{odd}
\end{array}
\right.
\label{eq:detailed}
\ee
where $i$ refers to the labelling in Fig.~\ref{fig:choices}. The wavefunctions
are illustrated in Fig.~\ref{fig:villain}. 
The pattern of phases is equivalent to that of the ordered groundstate of the fully-frustrated classical XY model\cite{villain}.

Since the condensed state is a superposition of states with different
$k_y$, it is a state with broken translational invariance.  The
difference in wavevectors is $\Delta k_y = k_y^A - k_y^B = \pi$, so
the new unit cell in the $y$-direction has size $\Delta y = 2$ (in
units of the lattice constant $a_y$).

It is useful to recall that the Hamiltonian at $\alpha=1/2$ is
time-reversal invariant, so all of its eigenvectors can be chosen
real. The fact that the condensed wavefunctions
(\ref{eq:detailed}) are imaginary shows that these break
time-reversal symmetry. The two states are time-reversed partners,
$\psi^c_\pm = \left(\psi^c_\mp\right)^*$.

The two states are the two (time-reversed) partners of a so-called
``staggered flux'' phase, a related version of which has been
discussed in the context of the cuprate superconductors.\cite{marston} This phase is characterised by circulating currents
around the plaquettes, which are arranged in a staggered
``checkerboard'' pattern.  It is straightforward to show that the
states $\psi^c_\pm$ do carry these staggered gauge-invariant
currents. 
We emphasize that the underlying Hamiltonian is both translationally
invariant and time-reversal invariant, so the emergence of these
staggered flux states is a result of the breaking of both of these
symmetries. 

The same condensed states have been shown to describe the groundstates
of a related Bose-Hubbard model.\cite{morais}  In constrast to the
  model that we study, in the model of Ref.~\onlinecite{morais}, the
  staggered flux $\alpha$ is applied directly in a checkerboard pattern.
  However, for staggered flux of $\alpha=1/2$ the two models coincide
  (up to a choice of gauge, as described below).

  The derivation of the states (\ref{eq:staggeredstates}) described
  above was presented in the weak-coupling limit, $nU \ll J$.
  However, note that the state has the special property that the
  density is uniform $|\psi_i^c|^2= \mbox{constant}$. 
Therefore, these
  condensates already minimize the interaction energy $\sum_i
  |\psi_i^c|^4$ at fixed average density. With increasing
  interaction strength $nU \sim J$, the condensate does not change.
  (We have confirmed this with extensive numerical simulations of the
  mean-field theory.\footnote{This was also found in the studies of Ref.~\protect\onlinecite{duriclee}. (D.K.K. Lee, private communication)})
  %\cite{footnotederekconfirm}) 
  Even in the regime $U\gg J$ where
  Gross-Pitaevskii theory becomes unreliable owing to the suppression
  of number fluctuations and a Gutzwiller mean-field theory is
  required, the condensate wavefunction remains the same.

\subsubsection{The case $\alpha =0.389$}
\label{seq:alpha_0.389}

\label{sec:3fold}

Choosing $\alpha =
\frac{1}{\pi}\arccos\left(\frac{1}{2}\sqrt{\frac{1}{2}(9-\sqrt{65})}\right)
\simeq 0.389$ leads to the situation in which $\Delta k_y = 2\pi/3$.
Then, a state of the form (\ref{eq:linear}) with non-zero $A$ and $B$
breaks translational symmetry with a period of $\Delta y = 3$ in the
$y$-direction.

Minimizing the interaction energy over states of the form
(\ref{eq:linear}) leads to the conclusion that the optimum condensate
wavefunction is the superposition
\be 
\psi^c_\pm = \frac{1}{\sqrt{2}}\left[ \psi_{k_A} + e^{i\chi}
  \psi_{k_B}\right].
\label{eq:d3}
\ee
Thus, the condensate does break translational symmetry, with the expected period
of $\Delta y = 3$. Unlike the (special) case of $\alpha=1/2$, for these
functions the density is not uniform, so spatial patterns of the
density show this periodicity.

One interesting feature of this mean-field result is that the
interaction energy is independent of the phase $\chi$ appearing in
(\ref{eq:d3}). Thus, there is an infinite set of condensed groundstates, which
are related by the different choices of the phase $\chi$.  In a continuum
model, this infinite degeneracy would correspond to the broken continuous
translational symmetry, the different choices of $\chi$ denoting translations
of the same state. Here the model is explicitly defined on a lattice, so there
is no continuous translational symmetry. Since the
wavefunction is defined only on lattice sites, the states with different
values of $\chi$ are not just related by simple translations. 
Nevertheless, for weak interactions
$nU \ll J$, we find that, surprisingly, there emerges a continuous degeneracy, beyond the expected threefold degeneracy from translations. 
This gives an additional ``Goldstone'' mode, associated with slow
spatial variations of $\chi$ as a function of position.
(The same feature arises for values of $\alpha$ for which $\Delta k_y = 2\pi/p$ with $p\neq 2$.)

Since this additional emergent degeneracy is not protected by a symmetry of
the Hamiltonian, one does not expect it to survive in general (e.g. to
stronger interactions or to quantum fluctuations). Indeed, we find that by
performing mean-field theory to stronger interaction strengths, the continuous
degeneracy is lost. Including energy corrections to order $(nU)^2/J$ we find
that the three cases $\chi = 0, \pm 2\pi/3$ are selected as the energy minima.
This gives rise to a set of three degenerate groundstates, with a unit cell of
size $\Delta y =3$ and all related by translations in the $y$ direction. The
Goldstone mode described above develops a gap.

\subsubsection{The case of general $\alpha$}

For typical cases of $\alpha$ in the range $\alpha_c < \alpha < 1/2$,
there are two minima in the single particle energy spectrum, but the
spacing $\Delta k_y$ is {\it incommensurate} with the underlying
lattice. In the weak coupling limit $nU \ll J$, one expects the
groundstate to remain incommensurate with the lattice, having a
continuous degeneracy (along the lines of that described by the phase
$\chi$ above). However, for sufficiently strong interactions, the
density wave will ``lock'' to the lattice, via an
incommensurate-commensurate phase
transition.\cite{pokrovskytalapov} 
Additionally, for staggered flux $\alpha \gtrsim \alpha_c$, near the bifurcation 
in Fig.~\ref{fig:minima}, the single-particle dispersion becomes very flat, and 
we expect a regime of large fluctuations where potential condensate groundstates 
will be strongly depleted.

\section{Numerical Methods}

In order to investigate the quantum groundstate of the frustrated
Bose-Hubbard model, we have performed large scale exact
diagonalization studies of the model (\ref{eq:hamiltonian}).  
We
define the system on a square lattice of $N_s = L_x\times L_y$ sites, and
impose periodic boundary conditions to minimize finite size effects.
For a total number of bosons, $N$, we therefore study a system with
mean density $n= N/(L_xL_y)$.

The possible geometries and system sizes are constrained by the form
of the gauge potential that is applied.  Furthermore, the plaquette
fluxes determine the translational symmetries of the Hamiltonian, and
hence the conserved momenta.  For staggered flux (\ref{eq:staggered}),
in the general case ($\alpha\neq 0,1/2$) the translational symmetries are 
those of the unit cell with size $2\times 1$ (in $x$ and $y$ directions).
Higher symmetry arises for the cases $\alpha=0$ and $\alpha=1/2$, for
which the flux is uniform; in the latter case, the (magnetic)
translational symmetries of many-particle systems follow from
Ref.~\onlinecite{kolread}.

By definition, the Hamiltonian commutes with the unitary
transformations that effect these translational symmetries. The energy
eigenstates are therefore also eigenstates of these translational
symmetry operators ({\it i.e.} eigenstates of conserved lattice
momentum). In any exact numerical calculation, the energy eigenvectors
will also be eigenstates of the conserved lattice momentum.  However,
as described above, in general the mean-field states break the
translational symmetry. Thus the mean-field states are not eigenstates
of momentum.  In order to make comparisons with the mean-field states,
and the possibility of condensation, one must allow for this breaking
of translational symmetry.

\subsection{Condensate Fraction}

Often the groundstates of (repulsive) interacting bosons can be
understood in terms of Bose-Einstein condensation.  Interactions
between the particles lead to depletion of the condensate. If the
effects of interactions are very strong, these may even drive a phase
transition from the condensed phase into non-condensed phases. It is
therefore very important to know if the groundstate remains condensed.
(If not, then the system may be described by a novel, uncondensed,
and possibly strongly correlated phase of matter.)

The condensate fraction is quantified by the general definition
introduced by Yang.\cite{yang}
From the many-particle groundstate $|\Psi_0\rangle$, one forms the single particle density operator
\be
\label{eq:sp}
\rho_{ij} = \langle \Psi_0 | \hat{a}^\dag_i\hat{a}_j|\Psi_0\rangle \,,
\ee
a Hermitian operator, the trace of which $\sum_i \rho_{ii}$ is the
mean (total) number of particles, $N$.  Then, one finds the
eigenvalues of $\rho_{ij}$.  For ``simple'' BECs,\cite{Leggett01} the
spectrum has one eigenvalue which is of order {$N$},\cite{yang} and
which is therefore much larger than all others for large $N$ (the
thermodynamic limit). Denoting this largest eigenvalue $\lambda_0$,
the condensate density $n_c$ and condensate fraction $x_c$ for
average density $n$ are defined by\cite{yang}
\be
\label{eq:xc}
 x_c \equiv \frac{n_c}{n} \equiv \frac{\lambda_0}{N}\,.
\ee  The eigenvector of $\rho_{ij}$ corresponding to the largest eigenvalue
 is the condensate wavefunction, 
$\psi_i^{0}$.

While this method is applicable in the simplest of situations, it gives
misleading results in cases where the groundstate of the system breaks a
symmetry of the Hamiltonian in the thermodynamic limit. In that case it is
well known that, for a finite sized system (in which the ground state is an
eigenstate of all symmetry operators), an analysis of the density matrix
states shows a ``fragmented'' condensate in which there is more than one
eigenvalue of order $N$. (See Ref.~\onlinecite{mueller:033612} for a discussion of
fragmentation in the context of cold atomic gases.) The origin of this
fragmentation, and its relationship to symmetry breaking, is well-understood
for simple model systems with no condensate depletion, for example for
condensation of $N$ bosons in two
orbitals.\cite{mueller:033612,dalibardparity}  From the practical point of
view of exact numerical calculations, it is important to have a prescription
for how to quantify the degree of condensation in general -- {\it i.e.} in cases
where there are many degrees of freedom and condensate depletion can be
significant.

\subsection{Condensate Fraction with Symmetry Breaking}
\label{subsec:symmetrybreaking}

We propose a method to determine, on the basis of numerical exact
diagonalization studies, the condensate fraction in cases where the
condensed state breaks a symmetry in the thermodynamic limit.  Given
the context of this paper, we focus on the case of translational
symmetry. The method, however, is very general: no specific knowledge
is required of the condensed state, or indeed of the symmetry that is
broken. These emerge directly from the numerical calculations in an
unbiased way.

The starting point is to determine the energy spectrum, in order to identify
if the groundstate may have a broken symmetry in the thermodynamic limit.  As
ever with numerical studies, one must study the spectrum with varying systems
size (up to as large a system size as can be achieved), in order to glean
information about the properties of the spectrum in the thermodynamic limit.
It is well-known that the signature of (translational) symmetry breaking in a
finite size calculation is the appearance of a set of quasi-degenerate energy
levels in the spectrum, with different eigenvalues of the conserved quantity
associated with the symmetry ({\it i.e.} momentum in the case of translational
symmetry breaking).  In the context of cold atomic gases, this has been
illustrated for spin-rotational invariance,\cite{LawPB98,hospins}
translational and rotational symmetry breaking,\cite{cwg,advances,zhaimtm} and
parity.\cite{parkeparity,dalibardparity} In the thermodynamic limit, these
states become degenerate, and it becomes valid to superpose the states
(e.g. as selected by an arbitrary weak symmetry breaking perturbation). Any
superposition of all these states forms a groundstate which has broken
symmetry. Before superposition, each of the states (eigenstates of the
symmetry operators) can be viewed as fragmented
condensates.\cite{mueller:033612}

Based on the results of these calculations of the energy spectrum, one
can look to see if, in the thermodynamic limit, several states are
tending to become degenerate.  The emergence of this degeneracy
appears when the system size is sufficiently large; if the degeneracy
is not well resolved, then this suggests that the numerical
calculations are not on a sufficiently large system size to be
conclusive.  In many cases, an emergent quasi-degeneracy can be very
convincingly
established.\cite{cwg,advances,parkeparity,dalibardparity} In many
practical cases of interest where the degeneracy is not fully
established, it can still be of value to make the hypothesis that a
small number of low-energy states will be degenerate in the
thermodynamic limit, and to test if this hypothesis is borne out by a
high condensate fraction.  

Suppose that an analysis of the energy spectrum suggests that there
are $D$ such states, $|\Psi_0^{\mu}\rangle$, with $\mu=1,2,\ldots,D$,
which tend towards degeneracy in the thermodynamic limit.  We assume
that these $D$ states can be distinguished by eigenvalues of symmetry
operators (e.g. no two have the same momenta). 
In order to investigate the possibility of simple BEC in a
broken symmetry state, we propose that one forms the superposition
state
\be
|\Psi^{c}_0\rangle \equiv \sum_{\mu=1}^D c_\mu |\Psi_0^\mu\rangle
\label{eq:psic}
\ee
which depends on the $D$ complex amplitudes $c_\mu$. Then, for this
superposition state -- which is not an eigenstate of the
(translational) symmetry -- one should determine the single particle
density matrix (\ref{eq:sp}) and find the condensate fraction
(\ref{eq:xc}), each of which are functions of the parameters $c_\mu$.
We define the condensate fraction of the broken symmetry state by in
terms of the optimal choice 
\be 
X_c \equiv \max_{c_\mu}\left[
  x_c(c_\mu) \right]\,.
\ee
The corresponding optimizing coefficients $c_\mu$ define the associated
condensate wavefunction (\ref{eq:psic}).  Since this is a broken symmetry
state, in general there are $D$ sets of coefficients $c_\mu$ which give the
(same) maximum condensate fraction, and hence $D$ such condensed states.
These correspond to the $D$ broken symmetry states, and are related by
applications of the symmetry operations.

Below, we illustrate the application of this approach for the cases of
the Bose-Hubbard model with staggered flux, at $\alpha=1/2$ (fully
frustrated) where $D=2$, and at $\alpha=0.389$ where $D=3$.

\subsection{Unfrustrated Bose-Hubbard Model}

As a warm-up, and to test the quantitative validity of exact
diagonalization in determining the condensate fraction, we study the
Bose-Hubbard model in the absence of gauge fields (all plaquette
fluxes vanish, $n^a_\phi =0$). In this case, for non-integer particle
density, it is known that the groundstate is condensed,\cite{fisherbh}
and the condensate fraction has been established by detailed numerical
studies including quantum Monte Carlo,\cite{Hebert:2001p35}  
as well as in spin-wave theory.\cite{Bernardet:2002p34} For $n=1/2$, the
condensate fraction is $x_c = 1$ for $U/J=0$, and falls to $x_c \simeq
0.4$ in the hard-core limit $U/J\to \infty$.\cite{Bernardet:2002p34}

We have used ED results on systems of up to $L_x \times L_y =
  5 \times 6$ to determine the condensate fraction. Consistent with the lack
of symmetry breaking, in all cases the spectrum shows a clear
groundstate. (The broken gauge invariance of this state will appear in
a emergent quasi-degeneracy at different particle numbers.  We work at
fixed particle number.)  By forming the single particle density
operator, and finding its maximal eigenvalue, we find that for
hard-core bosons at $n=1/2$ the condensate fraction is $0.433(6)$.  The
favourable comparison with the Monte Carlo result illustrates that,
for this case, the system sizes amenable to ED are sufficiently large
to allow accurate quantitative determination of $x_c$.

\subsection{Fully Frustrated Bose-Hubbard Model, $\alpha =1/2$}

We now turn to the case of the fully frustrated Bose-Hubbard model,
$\alpha=1/2$, which is gauge equivalent to uniform plaquette fluxes
$n_\phi = 1/2$.  To analyse this case, we study system sizes for which
the translational symmetry\cite{kolread} is the largest, implying the
largest possible Brillouin zone for the conserved (many-particle)
momentum, and no degeneracy associated merely with the magnetic
translations.  (This is the ``preferred'' case of $d=1$ in the
terminology and notation of Ref.~\onlinecite{kolread}.)

\begin{figure}
\includegraphics[width=0.95\columnwidth]{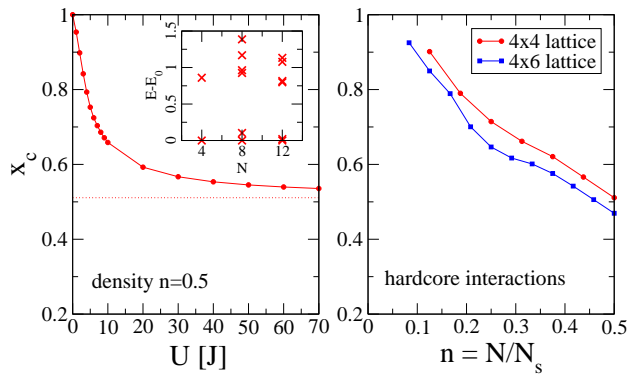}
\caption{Exact diagonalisation results for
  the condensate fraction $x_c$ for the maximally condensed state at
  $n_\phi=1/2$, {\it i.e.} incorporating translational symmetry breaking: as
  a function of the interaction strength $U$ (left); and as a function
  of particle density $n$ in the hard-core limit (right). In addition, the inset in
  the left frame shows the low-lying spectrum of the system for the case of hardcore
  interactions at half filling. These spectra clearly confirm the emergence of a twofold
  degenerate groundstate in large systems. At $N=12$, the splitting is barely discernable
  at the scale of the figure.
  }
\label{fig:fffraction}
\end{figure}

In these cases, where no \emph{many-body} degeneracy is expected, 
the groundstate in the non-interacting system still remains degenerate
due to the degeneracy of the \emph{single-particle} groundstate. This degeneracy 
is split due to interactions in the system. We find an emerging
two-fold quasi-degeneracy in the spectrum for sufficiently large
system sizes even in the case of hardcore interactions, 
as shown in the inset of Fig.~\ref{fig:fffraction} (left frame). Following the
procedure of \S\ref{subsec:symmetrybreaking}, we recognize this as a sign
of possible symmetry-breaking with $D=2$. We apply the prescription
in \S\ref{subsec:symmetrybreaking} to determine the maximal condensate
fraction.

In Fig.~\ref{fig:fffraction} we present the results of these
calculations of the (maximal) condensate fraction for the fully
frustrated Bose-Hubbard model, both as a function of $U/J$ for
$n=1/2$, and as a function of $n$ in the hard-core limit $U/J\to
\infty$. In all cases, we find that the groundstate appears to be fully condensed.
The smallest condensate fraction is for $n=1/2$ and $U/J\to \infty$.
Here, an extrapolation in $1/N$ to the thermodynamic limit yields
a condensate fraction of about $x_c \simeq 0.39$.
In the graphs of
Fig.~\ref{fig:fffraction}(b) there appears to be a reduction of
condensate fraction at about $n\simeq 0.25$.  We associate this with
the fact that, on the lattice, the Laughlin state of bosons would
appear for full frustration $n_\phi =1/2$ at density $n=1/2$.\cite{mollercooper-cf}
Extrapolation of the condensed fraction in this case yields $x_c \sim
0.48(2)$. Although the possibility of Laughlin correlations may act to
destabilize the condensed state, the groundstate at this density is a
condensed (superfluid) phase.
In all cases, the nature of the groundstate that we find  closely matches the
predictions of mean-field theory.  First, we can check that symmetry
breaking is of the same form. Furthermore, the condensate wavefunction that
we obtain is {\it exactly} that described above in Eq.~(\ref{eq:staggeredstates}).
As discussed previously, these states have uniform density, which makes them 
highly robust to details of the interaction potential.

The results provide the first evidence from exact diagonalizations
that, under all conditions, the groundstate of the fully frustrated
Bose-Hubbard model is condensed. Our model includes the
possibility of particle-number fluctuations, and thus goes beyond
previous studies using the picture of Josephson-junction arrays
that is based on phase fluctuations only.\cite{Polini:2005p392}
Furthermore, our results provide a quantitative measure of the condensed fraction.

It is interesting to compare the results to those that would be
obtained from a Gutwiller ansatz. In the hard-core limit, the
Gutwiller mean-field state has $x_c = 1-n$. 
Thus, at $n=1/2$ the Gutzwiller theory predicts a condensate fraction
of $0.5$. This is close to the value we obtain from exact
diagonalisation results, $\simeq 0.39$, indicating that Gutzwiller theory is
quantitatively fairly accurate in this case.
At $n=1/4$ the Gutzwiller theory predicts a condensate fraction of
$0.75$. The exact diagonalization result of $\simeq 0.48$, shows a large
quantitative discrepancy. As described above, we attribute this to the
competition introduced by Laughlin-like correlations which act to
destabilise the condensate.
Another interesting observation can be made by comparing the condensate
fractions for the frustrated Bose Hubbard model at $n_\phi=1/2$ with the
unfrustrated zero-field case. For half filling, the gauge-field reduces the
condensate fraction seen in our exact diagonalizations from about $x_c(n_\phi=0)=0.43$
to $x_c(n_\phi=1/2)=0.39$. For $n=1/4$ on the other hand, the reduction is more significant,
with $x_c(n_\phi=0)=0.66(1)$ being reduced to $x_c(n_\phi=1/2)=0.48$ in the presence
of the field.

\subsection{Staggered Flux Bose-Hubbard Model, $\alpha=0.389$}

This is the staggered flux value at which $\Delta k_y = 2\pi/3$, so we
expect a broken symmetry state with unit cell size $\Delta y =3$, and thus a
groundstate degeneracy of $D=3$ in the thermodynamic limit. As described
above, this is borne out in mean-field theory, at least for sufficiently
strong interactions.

\begin{figure}
\includegraphics[width=0.9\columnwidth]{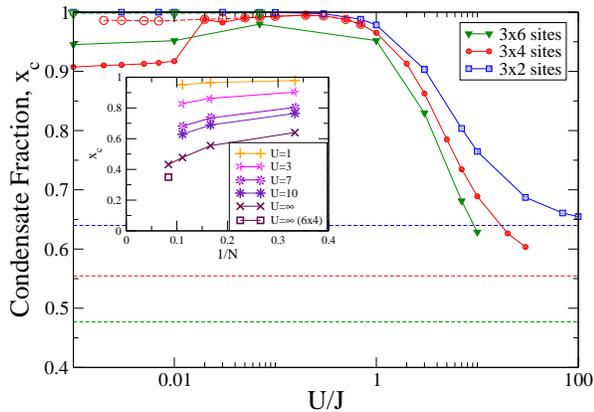}
\caption{ Exact diagonalisation results for
  the condensate fraction $x_c$ as a function of interaction strength $U/J$,
  for the staggered flux with $\alpha=0.389$ where the groundstate shows
  three-fold translational symmetry breaking. Results are at half filling,
  $n=1/2$, calculated for $N=6$ particles on a system of dimensions $L_x\times
  L_y=4\times 3$. Small symbols show optimisations over the lowest three
  eigenstates. For $U/J\lesssim 0.1$ a maximally condensed state requires to
  also include the low-lying Goldstone modes (large symbols; see main text). 
  The condensate fraction calculated in the hardcore limit is
  shown as a dashed line. Inset: Scaling of the condensate fraction with
  system size for different $U$. For the largest system with $N=12$ particles
  in $N_s=24$ sites, the two available lattice geometries yield significantly
  different $x_c$.}
\label{fig:3fold}
\end{figure}

The numerical results are consistent with these expectations. For large
interactions, a clear three-fold degeneracy appears in the groundstate.
Assuming $D=3$ for all interaction strengths leads to the condensate fraction
shown in Fig.~\ref{fig:3fold} for density $n=1/2$. 
For small $U/J$ and/or density $n$, the results of our analysis show that the 
groundstate is condensed in the manner predicted by mean-field theory. 
For the strongest interactions (hard-core interactions and $n=1/2$) finite-size 
effects remain significant, and it is difficult to be sure that extrapolation to the 
thermodynamic limit will leave a non-zero condensate fraction (see inset of Fig.~\ref{fig:3fold}).
In finite size systems, the condensed wavefunction obtained from the numerical 
procedure is in good qualitative agreement with the results of mean-field theory 
described above.  At very weak interactions, the presence of the Goldstone mode 
discussed in \S\ref{seq:alpha_0.389} is also visible. For $U/J\lesssim 0.1$, the
results in Fig.~\ref{fig:3fold} show a discontinuous drop in $x_c$. This occurs
when the splitting of the three-fold groundstate degeneracy (due to finite
size effects) is larger than the energy scale which mean-field
theory shows is required to ``lock'' the density wave to the underlying
lattice. Thus, this reduction in $x_c$ at small $U/J$ is a finite-size
effect. In this regime, we can recover a large condensate fraction, $x_c \simeq 1$, by including additional levels ($D>3$) to account for the  higher degree of symmetry breaking.

\section{Experimental Consequences}

The phases described above break translational symmetry of the
underlying lattice (with 2-fold and 3-fold symmetry breaking in the
cases discussed in detail in sections \ref{sec:alpha_0.5} and \ref{seq:alpha_0.389}).  
Following the usual expectations for
symmetry breaking states, in an experiment on a system with a large
number of atoms, one expects that very small perturbations (perhaps in
the state preparation) which break the perfect symmetry of the
underlying model will cause the system to select one of the broken
symmetry groundstates. (It is also possible that domains will form,
separated by domain walls that can be long-lived and survive as
metastable configurations.)

In general, one expects that this translational symmetry breaking will
appear in the real-space images of the system ({\it i.e.} if in situ imaging
on the scale of the lattice constant is possible). 
This is the case for the $3$-fold symmetry breaking described in
\S \ref{sec:3fold}. There, the real-space image of the particle density
will have spatial structure with a unit cell that has size $3\times
2$, and is therefore 3 times larger than the unit cell of the
underlying microscopic model. The three broken symmetry phases can be
distinguished by the three possible positions of this unit cell.
However, for $n_\phi=1/2$, where there is a 2-fold degeneracy of the
groundstate, the two broken symmetry phases {\it cannot} be
distinguished by the real-space image. Here, the groundstate is the
staggered flux phase.  This has both broken translational symmetry and
broken time reversal symmetry. However, it is invariant under the
combined action of translation and time reversal.  Since density is
time-reversal invariant, the state has uniform density.  Thus, one
cannot distinguish the symmetry breaking in real-space images.  (In
non-equilibrium situations where there can be domain walls separating
different phases, there may appear density inhomogeneities associated
with the domain walls.)  As we now discuss, there are ways to detect
these symmetry broken states in the expansion images.

\subsection{Expansion Images}

The nature of the groundstate can be probed by expansion imaging.  We
assume that all fields are released rapidly (compared to the subband
splitting of the lattice), and that the particles (of mass $M$) expand
ballistically -- without any potentials, or gauge fields and neglecting
further interactions -- according to the free-particle Hamiltonian,
denoted $\hat{H}_{\rm free}$. Then the expansion image after a time
$t$ is given by\cite{blochdz}
\be
n({\bm x}) = \left(M/\hbar t\right)^3|\tilde{w}({\bm k})|^2 G({\bm k})
\label{eq:exp}
\ee
where ${\bm k} = M{\bm x}/\hbar t$, $\tilde{w}({\bm k})$ is the Fourier transform of the Wannier state of the lowest Bloch band, and
\be 
\label{eq:ft}
G({\bm k}) = \frac{1}{N_s}\sum_{i,j} e^{i{\bm k}\cdot\left({\bm r}_i -{\bm
      r}_j\right)}\langle \hat{a}^\dag_i \hat{a}_j\rangle 
\ee
is the Fourier transform of the single particle density matrix. 
For states that are well-described as condensates ({\it i.e.} with small condensate
depletion), the density matrix is $\langle \hat{a}^\dag_i \hat{a}_j\rangle
\simeq (\psi^c_i)^*\psi^c_j$, where $\psi^c_i$ is the condensate wavefunction.
Thus, the expansion image directly provides (the Fourier transform of) this
condensate wavefunction.
As described in detail below in \S\ref{seq:MeasureCondensateFraction}, depletion leads to a reduction of the
amplitudes of the ``condensate'' peaks in the expansion image and to the
appearance of an additional incoherent background.

Compared to the usual Bose-Hubbard model, in the situation that we
consider here -- of an optically induced gauge potential on the
lattice -- there are three important new considerations concerning
expansion images.

$\bullet$ The first aspect relates to {\it gauge invariance}.  Consider making a change
of the vector potential in the Bose-Hubbard Hamiltonian
(\ref{eq:hamiltonian}), from $A_{ij}$ to
\be
A'_{ij} = A_{ij} + S_i - S_j
\label{eq:gauget}
\ee
where $S_i$ is any set of real numbers.  This ``gauge transformation''
leaves the fluxes (\ref{eq:flux}) unchanged, which are therefore
said to be ``gauge-invariant''. The new Hamiltonian (with $A'_{ij}$ in
place of $A_{ij}$) can be brought back to its original form
by introducing the operators
\be
\hat{a}'_i \equiv \hat{a}_i e^{-iS_i} \quad
\hat{a}'^{\dag}_i \equiv \hat{a}^\dag_i e^{iS_i} \,.
\label{eq:redef}
\ee
In this way, the Bose-Hubbard Hamiltonian adopts the same form as at the start
(again with gauge fields $A_{ij}$), but now with $\hat{a}'$ replacing
$\hat{a}$. Any property that is insensitive to the distinction between
$\hat{a}'$ and $\hat{a}$ in (\ref{eq:redef}) remains the same, and is
therefore gauge invariant.  Such quantities include the energy
spectrum, density response functions, etc.; indeed any observable of
the closed system described by (\ref{eq:hamiltonian}) is
gauge-invariant. All of these gauge-invariant properties depend only
on the gauge-invariant fluxes (\ref{eq:flux}).

An important point is that 
the expansion image is {\it not} gauge-invariant.
Under the transformation (\ref{eq:redef}) the single particle density
operator becomes
\be
\langle  \hat{a}'^{\dag}_i \hat{a}'_j \rangle = 
e^{i (S_i - S_j)}\langle  \hat{a}^{\dag}_i \hat{a}_j \rangle 
\ee
The gauge transformation affects the Fourier transform of the density
operator (\ref{eq:ft}), and therefore the expansion image
(\ref{eq:exp}).  There is no inconsistency with general principles of
gauge invariance.  As described above, prior to expansion, all
physical properties of the Bose-Hubbard Hamiltonian
(\ref{eq:hamiltonian}) are completely unchanged.  The essential point
is that the expansion image involves the evolution of the system under
the free-space Hamiltonian, $\hat{H}_{\rm free}$, and this Hamiltonian
is unchanged ({\it i.e.} the gauge transformation was not applied to it).
Indeed, if for the expansion imaging all optical dressing is switched
off, then this Hamiltonian is always in a fixed gauge with vanishing
vector potential. Provided the expansion image is taken under the
evolution of a Hamiltonian $\hat{H}_{\rm free}$ with vanishing gauge
potential, the expansion image measures the {\it canonical} momentum
distribution of the particles, and  therefore depends on the gauge
used in the Bose-Hubbard Hamiltonian $\hat{H}$ before expansion.

$\bullet$ The second consideration relates to the fact that, owing to the
optical dressing, {\it the atoms are in
 more than one internal state}.  In particular, for the schemes of
Refs.~\onlinecite{gerbier} and \onlinecite{JakschZoller}, the atoms on alternating sites along
the $x$ axis are, in turn, in the ground state, $|g\rangle$, and an excited
state, $|e\rangle$, of the atom.  Therefore, upon release of the cloud,
one has the possibility to study expansion images of several different
types.  The results described above (\ref{eq:exp},\ref{eq:ft}) apply
only if the image at time $t$ is formed in such a way that the
measurement does not distinguish between the different internal states
of the atom.  However, since the electronic states, $|g\rangle$ and
$|e\rangle$, are very different, it is also possible to perform
measurements which are state-specific: one can form the expansion
image of the $|g\rangle$ atoms, or of the $|e\rangle$ atoms. In these
cases, one should replace $G({\bm k})$ in (\ref{eq:exp}) by
\be
G_{g/e}({\bm k}) = \frac{1}{N_s}\sum_{i,j\in g/e} e^{i{\bm k}\cdot\left({\bm r}_i -{\bm
      r}_j\right)}\langle \hat{a}^\dag_i \hat{a}_j\rangle 
 \ee
where the change is that the sums should be over those sites $i,j$ on
which the atoms are of type $g$ or $e$.

$\bullet$ Finally, the third consideration -- although not restricted to systems of this
type -- arises naturally from the ability to address sites of type $g$ and $e$
separately by spectroscopic methods. This allows for much more interesting and
useful possibilities in the expansion imaging, including the possibility to
{\it imprint phase patterns prior to expansion}.  Specifically, immediately
prior to expansion, one can choose to drive (coherent) transfer of the atoms
from one state to an other. For example, using coherent laser fields of the
same type as used to provide the laser-assisted
tunneling,\cite{JakschZoller,gerbier} one can choose to transfer all atoms
(initially of both $e$ and $g$ type) into a given ``target'' state (a
superposition of $e$ and $g$) while adding a spatially dependent phase,
$S^{\rm exp}_i$.  Alternatively, the phases can be imprinted by site-selective
potentials $V_i$ applied for a short time $t$, leading to $S^{\rm exp}_i =
V_it/\hbar$.  In the language of the earlier discussion, immediately prior to expansion one
has effectively applied a gauge transformation to the initial wavefunction. The expansion
image follows from (\ref{eq:exp}) but now with
\be
\langle  \hat{a}'^{\dag}_i \hat{a}'_j \rangle = 
e^{i (S^{\rm exp}_i - S^{\rm exp}_j)}\langle  \hat{a}^{\dag}_i \hat{a}_j \rangle 
\ee
replacing the density operator in (\ref{eq:ft}). (Since we chose the
same target state for all atoms, they are indistinguishable in the
final image so all sites contibute.)  
This additional freedom is not restricted to dressed atomic systems of
the type we have described. Indeed, the application of a
spatially-varying potential $V({\bm r}$) over a time $t$ to a
one-component condensate will cause the local phase to wind by $V({\bm
  r})t/\hbar$. If the time $t$ is short compared to microscopic
timescales, this can be viewed as an instantaneous phase-imprinting
prior to expansion. Techniques of this kind could be used to tune out
the ``shearing'' in the expansion images of Ref.~\onlinecite{spielmanfield}.

\begin{figure*}
\includegraphics[width=1.8\columnwidth]{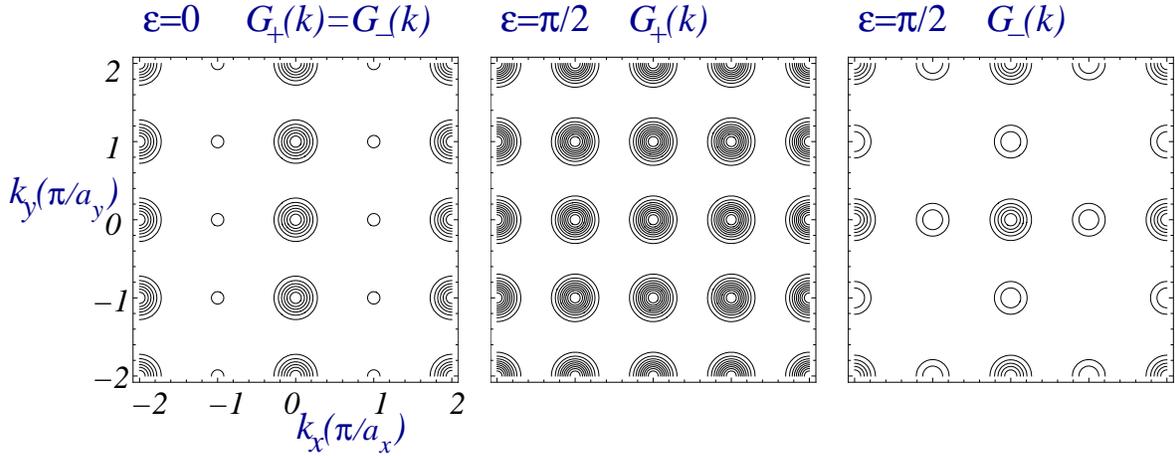}
\caption{The expansion image for the staggered flux phase, as determined by the structure factor as a function of $(k_x,k_y)$.
 (a) For the gauge field considered ($\epsilon=0$)
the two broken symmetry states have the same expansion image.
(b, c) With a phase imprinting before expansion, denoted by $\epsilon$, the two states have different expansion images.}
\label{fig:staggeredspectrum}
\end{figure*}

The additional freedom to imprint phases greatly enhances the
information that can be extracted from expansion images.  As we
illustrate below, it allows important information regarding the nature
of the phases to be extracted.

\subsubsection{Expansion Images for Full Frustration, $\alpha=1/2$.}
\label{sec:expansion_1_2}

As explained above, the groundstate at $n_\phi =1/2$ is the ``staggered flux''
phase, which is two-fold degenerate. 
In the gauge (\ref{eq:gauge}) we have 
been considering, the condensate wavefunctions 
are given by Eq.~(\ref{eq:detailed}), and
therefore have a unit cell of size $2\times 2$.
A straightforward calculation of $G^{\pm}({\bm k})$ shows that
the two broken symmetry states have the same
Fourier transform, $G^{+}({\bm k})=G^{-}({\bm k})$, illustrated in
Fig.~\ref{fig:staggeredspectrum} (a).  Therefore, the expansion
image cannot discriminate between whether the groundstate is in state
$\psi^c_+$
or in state $\psi^c_-$.
On the other hand, 
the expansion image can distinguish these states from the
expansion images of condensates formed from all particles condensed in
either one or the other of the two single-particle states.  These
individual single-particle states have the full translational symmetry
of the underlying system -- namely under $x\to x+2a_x$ and $y\to
y+a_y$ -- so the Fourier components of a condensate formed from either
one has peaks spaced by the reciprocal lattice vectors ${\bm
  K}_x=(\pi/a_x,0)$ and ${\bm K}_y=(0,2\pi/a_y)$. On the other hand,
the ``staggered flux phase'' has broken translational invariance in
the $y$ direction, being invariant only under the translations $y\to
y+2a_y$, so the Fourier components are spaced by the $(\pi/a_x,0)$ and
$(0,\pi/a_y)$.  The appearance of this smaller periodicity in the
$k_y$ direction is indicative of the broken spatial periodicity. 
\begin{figure}
\includegraphics[width=0.65\columnwidth]{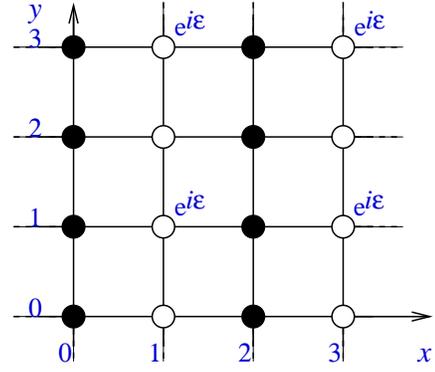}
\caption{A pattern of phases to imprint before expansion. For $\epsilon\neq 0$, this causes the
expansion images of the two staggered flux states to differ.}
\label{fig:phaseimprint}
\end{figure}

For the gauge used in this work, the expansion image does not
distinguish between the two different staggered flux states.  However,
in the gauge used in Ref.~\onlinecite{morais} these two states have very
different expansion images. Indeed, in the gauge of Ref.~\onlinecite{morais} 
one state has a condensate wavefunction with uniform density and phase (and 
therefore enjoys the full symmetry of the lattice), while the other has a phase
pattern with a $2\times 2$ unit cell; these give rise to very different
Fourier transforms and therefore expansion images.

As described above, the ability to locally address sites of the
lattice leads to the possibility to apply a spatial phase pattern
$e^{iS_i}$ to the system prior to expansion. This can be used to
discriminate between the two staggered flux states. Specifically, we
consider the case in which potentials or optical dressing is used to
imprint the phase pattern
\be
S_i = \epsilon\times{\rm mod}(x_i,2){\rm mod}(y_i,2)\,.
\ee
Thus, for the atoms on sites with $x_i= \mbox{odd}$
({\it i.e.} which are all of ``excited'' or all ``ground'' states in the
proposal of Ref.~\onlinecite{gerbier}), the phase of every other one
along the $y$-direction is advanced by $\epsilon$, as illustrated in
Fig.~\ref{fig:phaseimprint}. This can be achieved, for example, by
applying a state-selective potential with period $\Delta y = 2$.

\begin{figure}
\includegraphics[width=0.95\columnwidth]{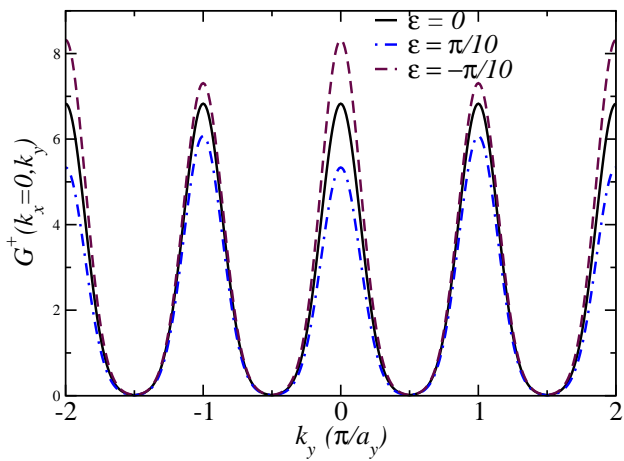}
\caption{The expansion image of the staggered flux state is very sensitive
  to $\epsilon\neq 0$, here shown for $\psi^c_+$. Even a phase of $\epsilon = \pi/10$ allows for notable
  change. Since $G_-({\bm k},\epsilon)=
  G_+({\bm k},-\epsilon)$, this change allows for a discrimination between $\psi^c_+$ and $\psi^c_-$.}
\label{fig:staggeredspectrum2}
\end{figure}

A calculation of the resulting expansion images shows that this phase pattern
causes the two staggered flux phases to have different expansion images,
$G_+({\bm k},\epsilon)\neq G_-({\bm k},\epsilon)$.  Owing to the relation
$\psi^c_+ = (\psi^c_-)^*$, it is straightforward to show that $G_-({\bm
  k},\epsilon)= G_+({\bm k},-\epsilon)$.  The results are illustrated for
$\epsilon = \pi/2$ in Fig.~\ref{fig:staggeredspectrum}(b) and (c).  There is a
clear distinction between the expanstion images of the two states: e.g. the signal
at $(k_x,k_y)=(1,1)$ is absent for $\psi_-$, while being strong for $\psi_+$.  It is
not, however, necessary to impose large phase difference $S_i$ to obtain large
effects.  The change in the spectrum increases linearly for non-zero
$\epsilon$. As shown in Fig.~\ref{fig:staggeredspectrum2} even very small
changes of phase $\epsilon$ can give rise to notable changes in the expansion
images.

\subsection{Measurement of the Condensate Fraction}
\label{seq:MeasureCondensateFraction}

The condensate fraction can be measured experimentally by analysing the expansion images of the lattice gas.
For a perfect condensate only few a coherent peaks are visible within the Brillouin zone.
Condensate depletion from strong interactions in the atomic gas results in the appearance of an additional
background, i.e. some of the density of particles is spread out in the Brillouin zone. 
Generally, we may write the density matrix of the depleted condensate as
\begin{equation}
\label{eq:depletedCondensateRho}
\rho_{ij} \equiv n_c (\psi^c_i)^* \psi^c_j + \delta\rho_{ij},
\end{equation}
where $n_c$ is the condensate density and $\psi^c_i$ is its wavefunction; this equation thus defines $\delta\rho_{ij}$.
Similarly, the amplitudes of the expansion image can be written as 
\begin{equation}
\label{eq:correctedGk}
G(\bk) \equiv n_c G_c(\bk) + \Delta G(\bk),
\end{equation}
where the coherent part $G_c(k)$ derives from the non-interacting condensate
\begin{equation}
G_c(\bk) = (1/N_s) \sum_{i,j}  (\psi^c_i)^* \psi^c_j \,e^{i\bk(\br_i-\br_j)}.
\end{equation}
This defines $\Delta G(k)$. Experimentally, one can only measure $G(\bk)$. 
In order to extract $n_c=n x_c$, one needs to make some assumption about $\Delta G(\bk)$. 
Numerically, we find that this background of the expansion image has some internal structure (and this data could be used to
build a more accurate model), but to a first approximation we may assume that it is homogeneous. 
This translates into making the simple assumption that $\Delta G(\bk)$ is independent of $\bk$. One then obtains
\begin{equation}
\label{eq:depletedCondensate}
G(\bk) \equiv n_c G_c(\bk) + (n-n_c),
\end{equation}
which satisfies the proper normalisation of the Fourier amplitudes $\sum_\bk G(\bk) = \sum_i \rho_{ii} = N$.

\begin{figure}
\includegraphics[width=0.8 \columnwidth]{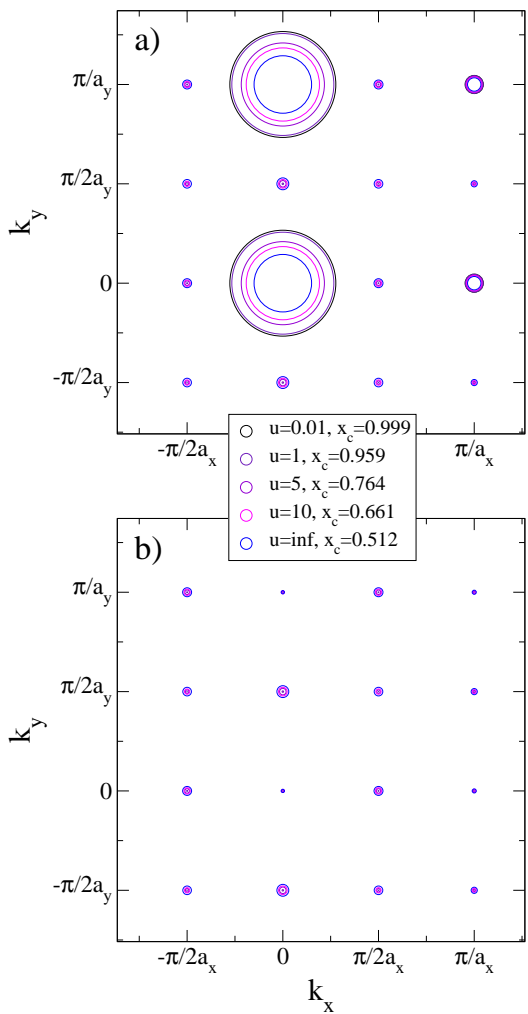}
\caption{a) Expansion images $G(\bk)$ of interacting Bose-Einstein condensates, showing the effect of condensate depletion. The
data show the evolution of the condensate at $n=n_\phi=1/2$ on a square lattice of size $L_x=L_y=4$ for weak to hardcore interactions. The depletion
of the coherent peaks is given approximately by the condensate fraction. b) The bottom panel shows the magnitude of the corrective
term $\Delta G(\bk)$ in Eq.~(\ref{eq:correctedGk}). This data shows some structure, in particular, the background contribution is smaller at 
the $k$-points where coherent peaks were present then it is elsewhere.}
\label{fig:depletion}
\end{figure}

Let us test the accuracy of this assumption for the example of a half filled lattice at $\alpha=1/2$ that was discussed in \S\ref{sec:expansion_1_2}.
To visualize the effect of condensate depletion, Fig.~\ref{fig:depletion} displays the evolution of the expansion
image for the case already displayed in Fig.~\ref{fig:staggeredspectrum}(a) for a weakly interacting gas.
Unlike in the single particle picture, calculations of the actual many-body wavefunction for the interacting system are limited
to finite size. Data in Fig.~\ref{fig:depletion} were obtained for a lattice of size $4\times 4$, so the `incoherent' background occurs at 
peaks spaced by $\Delta k_x = 2\pi/L_x = \pi/(2a_x)$ and $\Delta k_y = 2\pi/L_y = \pi/(2a_y)$.
The main features of the expansion image remain those of the pure condensate even with strong interactions, except
for the partial suppression of the coherent peaks that proceeds according to (\ref{eq:correctedGk}) with $\Delta G(\bk)$ given by small amplitudes.
The Brillouin zone can thus be partitioned into areas with peaks $\mathcal{A}_\text{P}$ and the remaining background area $\mathcal{A}_\text{BG}$.
The magnitude of this corrective term $\Delta G(\bk)$ is shown in Fig.~\ref{fig:staggeredspectrum}(b). Its spatial dependency is weak,
excepting the notably smaller correction within $\mathcal{A}_\text{P}$ as compared to the overall background $\mathcal{A}_\text{BG}$. 
Therefore, the average amplitude in the background 
\begin{equation}
\label{eq:DepletionProxy}
n-n_c \simeq \langle I(\mathcal{A}_\text{BG}) \rangle = \mathcal{A}_\text{BG}^{-1} \sum_{\bk\in\mathcal{A}_\text{BG}} G(\bk)
\end{equation}
is a proxy for the level of condensate depletion $(n-n_c)$.
Applied to the case with hardcore interactions in Fig.~\ref{fig:depletion}, we deduce a condensate fraction of $n_c=0.214$ from the intensity of the background, which 
should be compared to the exact value of $n_c=0.256$. This estimate is rather crude, as the background also contributes some signal within $\mathcal{A}_\text{P}$. 
A more accurate value is obtained allowing for such a contribution $\langle\delta I(\mathcal{A}_\text{P}) \rangle \equiv \kappa \langle I(\mathcal{A}_\text{BG})\rangle$,
where $\kappa$ is a `coherence' factor for the addition between the coherent condensate wavefunction and the incoherent background. The corrected estimate becomes
\begin{equation}
n-n_c =  \frac{\mathcal{A}_\text{BG} + \kappa \mathcal{A}_\text{P}}{\mathcal{A}_\text{BG}+\mathcal{A}_\text{P}}  \langle I(\mathcal{A}_\text{BG}) \rangle,
\end{equation}
which reduces to (\ref{eq:DepletionProxy}) in the limit of sharp coherence peaks $\mathcal{A}_\text{P}\to 0$. For the data in Fig.~\ref{fig:depletion}, we find exact values
of $\kappa$ in the range of 0.4 to 0.5. Assuming $\kappa = 0.45$, the exact condensate depletion is reproduced to within 1\% accuracy.

\section{Summary}

We have studied the groundstates of two-dimensional frustrated Bose-Hubbard
models. We have focused on the situation in which the imposed gauge fields
give rise to a pattern of staggered fluxes, of magnitude $\alpha$ and
of alternating sign along one of the principal axes. For $\alpha=1/2$ this
model is equivalent to the case of uniform flux $n_\phi=1/2$, which is the
``fully frustrated'' XY model with time-reversal symmetry.  We have shown
that, for $\alpha_c < \alpha < 1/2$, with $\alpha_c \approx 0.389$, 
the mean-field groundstate breaks
translational invariance, giving rise to a density wave pattern.  For $\alpha
=1/2$ the mean-field groundstate breaks both translational symmetry and
time-reversal symmetry, forming the staggered flux phase which has uniform
density but circulating gauge-invariant currents.

We have introduced a general numerical technique to detect broken
symmetry condensates in exact diagonalization studies.  Using this technique
we have shown that, for all cases studied, the Bose-Hubbard model with
staggered flux $\alpha$ is condensed.  We have obtained quantitative
determinations of the condensate fractions. In
particular, our results establish that the fully frustrated quantum XY model
is condensed at zero temperature, with a condensate fraction of $x_c\simeq
0.4$.  

The low-temperature condensed phases that appear in this system are of significant interest in
connection with their thermal phase transitions into the high-temperature
normal phase.  The groundstate of the fully frustrated system breaks both
$U(1)$ and $Z_2$ symmetries (owing to both Bose-Einstein condensation, and the
combination of translational symmetry breaking and time-reversal symmetry
breaking).  The transition to the high-temperature phase is interesting, combining the physics of
the Kosterlitz-Thouless transition [$U(1)$] with an Ising transition ($Z_2$),
and its properties have stimulated much theoretical debate.\cite{martinoli,Fazio2001235} The cold
atom system described here will allow this transition to be studied also in a highly quantum regime, with of order one particle per lattice site, 
that is inaccessible in frustrated Josephson junction arrays.\cite{Polini:2005p392}  Our results show that
the staggered flux model leads also to other cases that break $U(1)$ and $Z_3$
(or higher $Z_p$ symmetry depending on the flux $\alpha$). There is also the
possibility to study commensurate/incommensurate transitions driven by locking
of the density wave order to the underlying lattice.

We discussed in detail the experimental consequences of our results. In our
discussion, we have explained the meaning of gauge-invariance in ultracold
atom systems subject to optically induced gauge potentials: Expansion images
are gauge-dependent, and (provided gauge fields are absent during expansion)
measure the canonical momentum distribution.  Furthermore, we have shown how
the ability to imprint phase patterns prior to expansion (analogous to an
instantaneous change of gauge) can allow very useful additional information to
be extracted from expansion images.

\acknowledgments
GM gratefully acknowledges support from Trinity Hall Cambridge, and would like
to thank Nordita for their hospitality. NRC has been supported by
EPSRC Grant EP/F032773/1.

{\it Note added:}  After completing this work, we noticed a related preprint by Powell et al.,\cite{Powell:2010p391} which has
some overlap with our discussion of the case $\alpha=1/2$.

\bibliography{references}

\end{document}